# Towards an Approach for Analysing the Strategic Alignment of Software Requirements using Quantified Goal Graphs

Richard Ellis-Braithwaite[1]     Russell Lock[1]     Ray Dawson[1]     Badr Haque[2]
[1]Computer Science, Loughborough University                  [2]Rolls-Royce Plc.
Leicestershire, United Kingdom                                Derby, United Kingdom
{r.d.j.ellis-braithwaite@lboro.ac.uk, r.lock@lboro.ac.uk, r.j.dawson@lboro.ac.uk, badr.haque@rolls-royce.com}

*Abstract*—Analysing the strategic alignment of software requirements primarily provides assurance to stakeholders that the software-to-be will add value to the organisation. Additionally, such analysis can improve a requirement by disambiguating its purpose and value, thereby supporting validation and value-oriented decisions in requirements engineering processes, such as prioritisation, release planning, and trade-off analysis. We review current approaches that could enable such an analysis. We focus on Goal Oriented Requirements Engineering methodologies, since goal graphs are well suited for relating software goals to business goals. However, we argue that unless the extent of goal-goal contribution is quantified with verifiable metrics, goal graphs are not sufficient for demonstrating the strategic alignment of software requirements. Since the concept of goal contribution is predictive, what results is a forecast of the benefits of implementing software requirements. Thus, we explore how the description of the contribution relationship can be enriched with concepts such as uncertainty and confidence, non-linear causation, and utility. We introduce the approach using an example software project from Rolls-Royce.

## I. Introduction

This paper describes in more detail the concepts and the technique originally presented at the 7th International Conference on Software Engineering Advances [1]. It extends the work namely through a more comprehensive literature review, and the introduction of multi-point goal contribution.

The growth of the strategic importance of IT [2] necessitates the need to ensure that software to be developed or procured is aligned with the strategic business objectives of the organisation it will support [3]. Attaining this alignment is a non-trivial problem; firstly, decisions in the Requirements Engineering (RE) phase are some of the most complex in the software development or procurement lifecycle [4], and secondly, there is a gulf of understanding between business strategists and IS/IT engineers [5]. If alignment were trivial and easy, then it would not have been the "top ranking concern" of business executives for the last two decades [6], over 150 papers would not have been published on the topic [7], and perhaps there would be less software features implemented but never used (currently half of all features [8]).

Decisions made in the requirements phase greatly affect the value of the resulting software, e.g., in requirements prioritisation, the selection of the least important requirements allows costs to be cut by trading off the development of those requirements. The correctness of any such decision depends entirely on the availability of information about the choices available to the decision maker [9]. In this context, information about the value of a requirement, in particular, the causes and dependencies of value creation, is highly useful. Goal graphs are of great interest because they are well suited for visualising cause-effect, dependency, and hierarchical relationships between requirements [10].

This paper explores the suitability of goal graphs for demonstrating a software requirement's strategic alignment. Current Goal Oriented Requirements Engineering (GORE) approaches primarily take a qualitative or subjective approach to describing goal contribution, such as GRL's {--,-,+,++} or [-100,100] scores [11]. As a result, any strategic alignment proposed by the use of goal graphs is not specific, measurable, or testable. Proposed extensions by Van Lamsweerde [12] do not consider that a chain of linked goals may contain a variety of metrics that need to be translated in order to demonstrate strategic alignment. Additionally, the certainty, confidence, and credibility of the predicted contribution are not explored. A probabilistic layer for goal graphs is proposed in [13], which recognises that goals are often only partially satisfied by software requirements. However, the (often non-linear [14]) effects of the incomplete goal satisfaction on an organisation's various levels of business strategy are not explored. Furthermore, current methods do not consider how goal contribution scores are elicited [15], and how their calculation affects the credibility and accuracy of the claimed benefits. This paper attempts to demonstrate how the above problems can be addressed, thereby improving the applicability of goal graphs for the problem of analysing the strategic alignment of software requirements. By making assumptions about business value explicit, our approach complements Value Based Software Engineering (VBSE) [16].

We have developed and implemented our approach in partnership with an industrial partner (Rolls-Royce) to ensure its usability and utility in real world settings. We use examples in the context of a software project for a Business Unit (BU) responsible for part of a Gas Turbine (GT) engine, henceforth referred to as GT-BU. The software will automate geometry design and analysis for engine components, as well as for their manufacturing tools such as casting molds. Simply put, engineers will input the desired design parameters and the software will output the component's geometry. At the time of our involvement with the project, some high-level software requirements had already been elicited and defined according to the Volere template [17].

In Section II, we describe the problem that this paper addresses. Then, in Section III, we define and describe the essential terminology and concepts, while in Section IV, we present and evaluate the extent to which existing solutions address it. Section V presents our approach and tool in order to address the gaps outlined in Section IV. We conclude in Section VI with the paper's contributions and future work.

## II. THE PROBLEM

Stakeholders responsible for a software project's funding need to be able to demonstrate that the software they want to develop or purchase will be beneficial. Decision makers require granularity at the requirement level, rather than the project level, since individual software functions or qualities may significantly affect the benefit or cost of the software's development or procurement. Furthermore, stakeholders performing requirements engineering processes where the benefit of a requirement is questioned (e.g., in prioritisation, release planning, trade-off, etc.) need to know how benefit is defined by the stakeholders, and then how the requirements (and their alternatives) contribute to it.

As an example of the problem that this paper examines, we introduce the following high-level software requirement taken from our example project: "While operating in an analysis solution domain and when demanded, the system shall run analysis models". The rationale attached to this requirement is "So that structural integrity analysis models can be solved as part of an automated process". Although the rationale hints at automation, the requirement's benefit to the business and the potential for alignment with business strategy are unclear. In order to understand the latter (i.e., the alignment with business strategy), the extent to which the organisation wants to reduce the problems related to manual structural integrity analysis needs to be understood (i.e., its business objectives). In order to understand the former (i.e., the business value), the extent of the requirement's contribution to the problem to be solved needs to be made clear, e.g., the extent that automation is likely to solve the problems related to manual structural integrity analysis. For example, if the manual process costs the business in terms of employee time and computing time, how much time is consumed, and at what cost? Then finally, to what extent will the software requirement's successful (or partially successful) implementation reduce the time consumption and cost?

To paraphrase Jackson & Zave [18], for every stated benefit (or an answer to "why?"), there is always a discoverable super benefit (i.e., benefit that arises from that benefit). For example, the slow and human resource intensive process may constrain the designer's ability to innovate (by not being able to analyse new design ideas), which may ultimately harm the organisation's competitive advantage. Many levels of benefit follow a requirement's implementation. Each level provides the possibility of contributing to a business objective at a different level of the organisation. There are arguably more levels of benefit than it would be sensible to express within a requirement, since several requirements may achieve the same benefit, but their contribution will vary.

## III. BACKGROUND TERMINOLOGY

### A. Software Requirements

In 1977, Ross and Schoman stated that software requirements "must say why a system is needed, based on current or foreseen conditions" as well as "what system features will serve and satisfy this context" [19]. Robertson & Robertson later expanded the concept of a "feature" by defining a requirement as: "something that a product must do or a quality that the product must have" [17], otherwise known as functional and non-functional requirements, respectively. It is worth noting here that, according to the "what, not how" [20] paradigm, software requirements are often incorrectly specified in practice (i.e., they often describe *how* features should work, rather than *what* features should be implemented). Consequently, implementation bias may occur, unnecessarily constraining the design space. Practitioners are not entirely to blame however, since the *what* and *how* separation is confusing. This is because a requirement describes both concepts at the same time, but at different levels of abstraction. For example, "print a report" is what the system should do, but also how the system should "make the report portable" – which again, is what the system should do, but also how the system should "make reports shareable". The *how* and *why* aspects of a *what* statement are simply shifts in the level of the statement's abstraction (down, and up, respectively).

Popular requirements engineering templates (e.g., Volere [17] and IEEE Std. 830-1998 [21]) and meta models (e.g., SysML [22] and the Core Metamodel [23]) tend not to focus on the *why* aspect, typically addressing it by stipulating that rationale be attached to a requirement. However, rationale is not always an adequate description of why the requirement is valuable. If only one *why* question is asked about the requirement then the rationale can still be distant from the true problem to be solved (i.e., the essence of the requirement), and it may be defined without consideration of its wider implications. A stakeholder's "line of sight" (i.e., the ability to relate low-level requirements to high-level business goals), and thus, the ability to determine the value of a requirement, depends on their ability to find answers to enough recursive *why* questions. Anecdotally, empirical studies at Toyota determined that the typical number of *why* questions required to reach the root cause of a problem is five (thus spawning the "five whys" method popularised by Six Sigma) [24].

### B. Strategic Alignment

Singh and Woo define business-IT strategic alignment as "the synergy between strategic business goals and IT goals" [7]. In the IT context, Van Lamsweerde defines the term "goal" as a prescriptive, optative statement (i.e., desired future state) about an objective that the system hopes to achieve [25]. In the business context, a goal is defined as an abstract indication of "what must be satisfied on a continuing basis to effectively attain the vision of the business" [26]. In order to relate the goals of the system to those of the business, an integrated definition of the terms used by business strategists and software developers is required. Furthermore, the first definition does not make "objective" distinct from "goal". The Object Management Group (OMG) defines such terms in its Business Motivation Model (BMM) [26]. There, an objective is defined as a specific "statement of an attainable, time-targeted, and measurable target that the enterprise seeks to meet in order to achieve its goals". According to the definitions of goals and objectives provided by the BMM, the difference between a goal and an objective lies in the goal's hardness (i.e., whether the goal's satisfaction can be determined). Therefore, from now on, we use the terms "hard goal" and "objective" interchangeably.

Finally, the BMM defines that the performance of a business strategy (means) is measured by the business objectives (ends) that the strategy supports [26]. Thus, the extent to which a software requirement aligns to business strategy depends on the extent to which the requirement contributes to the satisfaction of the strategy's business objectives. Attempting to demonstrate a requirement's strategic alignment to soft goals (e.g., "maximise profit") would be inappropriate, since it would not be possible to describe the extent of the requirement's contribution to the goal. Therefore, when demonstrating strategic alignment, requirements should be related to objectives rather than goals.

## IV. Related Work

The following areas of research are related to analysing the strategic alignment of software requirements: (A) Value Based Software Engineering, (B) Goal Oriented Requirements Engineering, (C) Strategic Alignment Approaches, (D) Quantitative Requirements and Metrics, and (E) Requirements Traceability Approaches.

### A. Value Based Software Engineering (VBSE)

The VBSE agenda is motivated by observations that most software projects fail because they do not deliver stakeholder value, yet, much software engineering practice is done in a value-neutral setting (e.g., where project cost and schedule is tracked rather than value) [27]. Value Based Requirements Engineering (VBRE) takes the economic value of software systems into perspective through activities such as stakeholder identification, business case analysis, requirements prioritisation, and requirements negotiation [28]. The primary VBRE activities are Business Case Analysis (BCA) and Benefits Realisation Analysis (BRA) [16]. Other VBRE activities such as prioritisation are considered secondary to these, since they depend on benefit estimation [29].

In its simplest form, BCA involves calculating a system's Return on Investment (ROI) - which is the estimated financial gain versus cost, defined in present value. While simple in definition, accurately calculating ROI is complex, since the validity of any concise financial figure depends on assumptions holding true, e.g., that independent variables remain within expected intervals (e.g., time saved is between [x,y]). Estimating benefit involves further intricacies such as uncertainty and the translation of qualitative variables (e.g., the software user's happiness) to quantitative benefits (e.g., sales revenue) - none of which are made explicit by BCA. An advancement from BCA in descriptiveness, $e^3$value modelling seeks to understand the economic value of a system by mapping value exchanges between actors, ultimately leading to financial analysis such as discounted cash flow [30]. However, it does not address how economic value creation is linked to requirements, nor are links between value creation and business strategy attempted.

BRA's fundamental concept is the Results Chain [2], which visually demonstrates traceability between an initiative (i.e., a new software system) and its outcomes (i.e., benefits) using a directed graph, where nodes represent initiatives, outcomes, and assumptions, while edges represent contribution links. BRA's contribution links allow one initiative to spawn multiple outcomes, but the links are not quantitative, e.g., outcome: "reduced time to deliver product" can contribute to outcome: "increased sales" if assumption: "delivery time is an important buying criterion" holds true – but the quantitative relationship between "delivery time" and "sales increase" is not explored. This is problematic when outcomes are business objectives, since their satisfaction depends on the extent that they are contributed to, e.g., in the case of a cost reduction objective, the primary concern is the amount of reduction that is contributed by the actions.

In summary, neither BCA nor BRA estimates the benefit of individual requirements, but rather for whole systems. A similar criticism also applies to the majority of requirements prioritisation methods, as a systematic literature review "found no methods which estimate benefit for [primary] individual requirements" [29], nor were any found which derive the benefits of secondary requirements from their contribution to primary requirements. In this context, primary refers to stakeholder requirements or business objectives while secondary refers to those derived from the primary requirements (e.g., a refined functional requirement).

### B. Goal Oriented Requirements Engineering (GORE)

GORE seeks to provide answers to "why?" software functionality should exist. The most well-known GORE methodologies include KAOS [31], i* [32] and GRL [33]. Such methodologies produce goal graphs whereby goals at a high level represent the end state that should be achieved and lower level goals represent possible means to that end. A goal graph is traversed upwards in order to understand why a goal should be satisfied, and downwards to understand how that goal could be satisfied. In this context, a requirement is a low level goal where one agent (e.g., a human or a machine) is responsible for its satisfaction. Other related concepts such as resources, beliefs and obstacles are typically related to goals to describe what a goal's satisfaction requires, while agents indicate who is responsible for, dependent on, or wishes for a goal's satisfaction. Relationships between goals are typically represented by means-end links, where optional AND/OR constraints represent alternative options for satisficing a goal. Contribution links are enhanced means-end links, in that they describe the extent to which a goal contributes to the achievement of another. However, "extent" is usually defined in terms of sufficiency and necessity (logic), not as in the quantitative extent of the contribution [34].

#### i. Goal-Goal Contribution Links

Contribution links are usually annotated with a score or a weight to represent the degree of contribution made by the goal. Three approaches for applying scores to contribution

links in goals graphs are described by Van Lamsweerde [12]:
1. Subjective qualitative scores e.g., --, -, +, ++.
2. Subjective quantified scores e.g., -100 to 100.
3. Objective gauge variable (i.e., a measured quantity predicted to be increased, reduced, etc.).

After evaluating the above approaches, Van Lamsweerde concludes that the specification of contribution scores with objective gauge variables (the third option) is the most appropriate for deciding among alternatives, due to its verifiability and rooting in observable phenomena. Of course, the subjective approaches are no doubt quicker to use, but their sole use risks misunderstanding the actual contribution mechanism. A comprehensive comparison of the qualitative contribution reasoning techniques can be found in [35].

Just as objective contribution scoring adds rigour and testability to the task of deciding between alternatives, the same applies to the task of demonstrating alignment to business objectives. Thus, contribution scores should be quantified in terms of the contribution likely to be made to the objective. Our rationale is that, by definition, objectives are quantitatively prescribed, and reasoning qualitatively about degrees of satisfaction of a quantitative target is highly ambiguous. Additionally, this will allow the contribution scores to be verifiable so that they (as predictions) may be later compared to actual results in the evaluation stage of the project. It must be noted here, however, that this option is not without its disadvantages - empirical studies in requirements prioritisation show that practitioners find providing subjective data far easier than objective data [36]. A parallel can be drawn here to the decision analysis field, where inferior (i.e., qualitative) processes have found favour with decision makers because "they do not force you to think very hard" [9].

Van Lamsweerde goes on to explain how alternative goal (i.e., requirement) options can be evaluated by predicting contributions made by goals to soft goals (which define the qualities to be used for comparison) [12]. However, in the prescribed approach, the relationship between the soft goals and the predicted benefit to be gained by their achievement is not made explicit. In other words, the contribution scores are not abstracted to different levels of benefit such that they may eventually relate to business objectives. Each of these potential benefit abstractions require that the metrics used to measure contribution (and satisfaction) are translated (e.g., from time saved to money saved). Furthermore, the expected values allocated to the objective gauge variables are single-point representations of inconstant and variable phenomena. For example, when estimating the number of interactions required to "arrange a meeting via email" – an alternative requirement option taken from the paper's meeting scheduling system example – a single number does not describe the possible variance, or how that variance can affect the desired end. This is important for predicting strategic alignment, since variance in a requirement's satisfaction is likely [13], and it will affect the satisfaction of the related business objective(s).

GORE approaches typically describe a goal's benefit relative to other goals with an importance or weight attribute [12], where importance is a qualitative label (e.g., high, medium, low) and weight is a percentage (where the total of all assigned weights is 100%). Both of these approaches are ambiguous and subjective, and are not traceable to observable benefits, e.g., alignment with business objectives.

A probabilistic layer for quantified goal graphs is proposed in [13] to represent the variance of goal contributions in terms of Probability Density Functions (PDFs). However, effects of the variance on the satisfaction of high-level goals, or business objectives, are not described. To use the example provided in the paper, the effects of an ambulance arriving at a scene within 8, 14, or 16 minutes (i.e., satisfaction of the target exceedingly, completely, or partially) are not described in the context of the benefits of doing so – i.e., to what extent will some problem(s) be affected given these possible goal satisfaction levels. If this is not explored, then it might be that there are no significant benefits to be gained at certain intervals of goal satisfaction levels (note that this point is more significant for non-life-threating systems). Thus, if a goal is defined with a specific target (e.g., target arrival time) in mind, without the rationale for doing so explored as further goal abstractions, then satisfying that goal may not be worthwhile - "wrong decisions may be taken if they are based on wrong objectives" [13]. Furthermore, probabilistic approaches have limited applications (PDFs are not often available and are time consuming to construct), and do not capture stakeholder "attitude, preference and likings" [15].

*ii. The Goal-oriented Requirements Language (GRL)*

Given the choice of GORE methodologies, we chose to focus on and adopt GRL [33] for the following reasons:
1. GRL's diagrammatic notation is well known within the RE community (since it originates from i*) [33].
2. i* (GRL's primary component) has been shown to be the most suitable for modelling Information System (IS) strategic alignment according to the strategy map concept (GRL not included in review) [37].
3. GRL has an ontology describing its modelling concepts (where others are described informally) [34].
4. GRL was recently made an international standard through ITU specification Z.151 [11].

GRL integrates the core concepts of i* and the NFR Framework [33] (where i* inherits the qualitative goal contribution mechanism from NFR [32]), but GRL adds to i* the capability to express contributions quantitatively. Thus, goal contributions in GRL can be specified with either subjective numeric scores (interval [-100,100]) or qualitative labels (one of {--,-,+,++}) [33], i.e., the first and second options outlined in Van Lamsweerde's paper [12]. For example, a time reduction goal might contribute to an overall-cost saving goal with a contribution weight of 67 out of 100 with positive polarity (+). Such a contribution score is untestable and not grounded by observable phenomena. Moreover it is not refutable, which, according to Jackson [38], means that the description is inadequate because no one can dispute it. The only way such scales could be testable is if the goals were specified with fit criteria (e.g., a cost to be saved), which mapped to the scale, e.g., that they implied percentage satisfaction (which they do not). In which case, a 50/100 contribution might imply that 50% of a £20,000 annual cost saving will be achieved. However, this is only applicable for goals whose satisfaction upper bound is 100% (since the scale's upper bound is 100), which is not the case for goals involving increases (e.g., where a mean's contribution to an end exceeds the end's target level).

Recently, the jUCMNav tool allowed goals to relate to Key Performance Indicators (KPIs) in GRL, in order to map a goal's satisfaction value to real world numbers [39]. However, subjectivity still exists in goal chains (i.e., >1 link), since KPIs do not affect the way in which goal contribution is specified further up the goal chain (i.e., as low-level benefits are translated to high-level business objectives, e.g., converting time to cost). Also, the interaction between KPIs is not considered, e.g., composition via hierarchy or non-linear causation. Since the publication of our original work, Horkoff et al. have improved GRL's integration with indicators to consider the hierarchy of KPIs alongside a goal model [40]. However, their approach is concerned with improving Business Intelligence (BI), rather than aligning software requirements to strategic business goals. Thus, several areas are still lacking when applied to our problem. Means are not distinguished from ends (i.e., business objectives and software requirements), making it difficult to know which sets of goals should be aligned, or how those goals should be defined or organised differently. Also, stakeholder utility and confidence through the range of possible goal satisfaction levels (i.e., KPIs in the approach) is not specified – making it hard to know the effects of partial requirement satisfaction, or the credibility of the estimated alignment. Furthermore, non-linear relationships in the associated KPI hierarchy (i.e., diminishing returns in achieving an objective) are not amenable to algebraic description [14] (i.e., "business formulae", as termed in the approach) – making their definition and communication difficult. Finally, potential fluctuation or uncertainty (i.e., the range of possibilities) in goal contribution is not described, as is done with *usage profiles* in [41].

As an additional concern, a contribution link is underpinned by assumptions which can either make or break the satisfaction of the end goal. For example, a reduction in task time will only reduce costs if associated costs are actually cut (e.g., by billing work to a different project, or through redundancies). GRL's belief elements (otherwise known as "argumentation goals") could be used to express such assumptions in order to provide an integrated view, despite their inferiority in richness to satisfaction arguments [42]. However, in the case of this particular assumption, it seems more semantically appropriate to model it as a necessary action for the end-outcome, just as the BRA's Results Chain [2] does.

### C. Strategic Alignment Approaches

The Balanced Scorecard and Strategy Maps (SMs) approaches [43] offer guidance on formulating and relating business goals to each other under four perspectives: financial, customer, internal processes, learning and growth. In order to improve traceability between these perspectives, SMi* combines SMs with i* goal models [44]. While this approach does not directly relate to software requirements, goals could be categorised by the four perspectives to ensure coverage.

The most suitable framework for relating software requirements to business strategy is B-SCP [45], due to its tight integration with the OMG's BMM and the explicit treatment of business strategy that this affords [7]. B-SCP decomposes business strategy towards organisational IT requirements through the various levels of the BMM (i.e., the vision, mission, objective, etc.). However, B-SCP cannot show the extent to which a requirement satisfies a strategy, since no contribution strengths are assigned to links between requirements and the strategy's objectives. Moreover, B-SCP's methodology refines business strategy top-down towards IT requirements, which means that completeness of the model is dependent on the completeness of the business strategy, i.e., there is no opportunity to refine software functionality upwards to propose new business strategy. Additionally, B-SCP does not consider rich GORE concepts (e.g., AND/OR, actors), as found in GRL.

### D. Quantitative Requirements and Metrics

The contribution that a requirement's implementation makes to a business objective depends on the extent of the requirement's satisfaction (i.e., partially or completely). In order to understand the extent of a requirement's satisfaction, the desired outcome of the requirement must first be made explicit. Although its practicality is debated [46], it is considered best practice to describe a requirement's desired outcome using quantitative measures [47]. In [48], Gilb describes the steps that requirements quantification should entail. Firstly, the desired level of achievement should be elicited. Then secondly, the capabilities of the various alternative design solutions should be estimated against that desired level. Finally, the delivered solution should be continuously measured against that desired level. Unfortunately, these steps are rarely followed in practice [47], [48].

As a result of a career training practitioners to quantify requirements, Gilb concludes that there are two main obstacles to quantifying requirements [48]. Primarily, practitioners find defining quantitative scales of measure for a requirement difficult, often believing that it is impossible to quantify all requirements due to their sometimes qualitative nature (guidance on doing so can be found in [51]). Secondly, practitioners encounter difficulty in finding ways of measuring numeric qualities of software which are practical to use (i.e., meters in Planguage), and at the same time, measure real qualities. Besides, even if a requirement is quantified, its quality is not necessarily improved; a related survey revealed that precisely quantifying requirements can lead to long project delays and increased costs if the quantifications are unrealistic [49]. This is problematic, since it is not straightforward to determine what is realistic with current technology in order to set the desired level of achievement. Despite the difficulties in expressing a requirement's fit criterion quantitatively, qualitative descriptions (e.g., "good uptime") are too ambiguous to be useful – in both trying to achieve that requirement, and in analysing the effect of its implementation on the (quantitatively defined) business objectives. The only caveat to this is that qualitative terms such as "good" can be suitable if they have been defined as fuzzy numbers [50].

The Volere [17] template stipulates that a "Fit Criterion" be attached to a requirement in order to make its satisfaction empirically testable (i.e., the first step of Gilb's requirement quantification steps). Planguage [51] similarly provides a template for describing how a requirement's satisfaction should be tested, and what the result of the test should be. Planguage's fit criteria are more descriptive than Volere's, since multiple levels of quantitative fit criterion are specified, e.g., for what must be achieved (minimum), what is planned to be achieved (likely), what is wished to be achieved (best case), and what has been achieved in the past (benchmark).

GQM+Strategies [52] was developed to extend the Goal Question Metrics methodology by providing explicit support for the traceability of software metrics measurement effort at the project level (e.g., measuring the impact that pair programming has on quality) to goals at the business level (e.g., increasing the software user's satisfaction). In [53], the approach is used to show the alignment of software project goals to high-level business goals using a 2d matrix. The approach's main benefits are that goals are defined quantitatively using a tried and tested metrics template, and, that assumptions which underpin goal to goal contributions are made explicit, much like GRL's belief element allows for. However, the approach focuses on decisions at the project level, rather than the requirements level (i.e., which projects, rather than requirements, should be implemented?), so is not directly applicable to the problem – a large and variable number of goal abstractions can be required to link a requirement (a means) to a project goal (an end). Additionally, the approach falls short in areas similar to the other methodologies reviewed. Firstly, when a link exists between two goals, the effects of the first goal's satisfaction on the second goal are not explored. Thus, although each goal has a target satisfaction level (e.g., 5% profit increase), the predicted contributions that its child goals (e.g., software requirements) make toward it are not represented (along with forecast related information such as confidence, evidence, stakeholder agreement, etc.). Therefore, although GQM+Strategies achieves traceability between project goals and business goals, it is not possible to analyse the *extent* of the strategic alignment of software requirements, since, as aforementioned, requirements often partially satisfy goals [13], i.e., the effect of a requirement's partial satisfaction is not described in the context of business objectives. Finally, the approach lacks concepts found in GORE which contextualise goals and support decision analysis (e.g., actors, obstacles, AND/OR links).

### E. Requirements Traceability Approaches

Several approaches exist which allow means to be traced to ends, typically by constructing a 2d comparison matrix where rows list means and columns list ends. Such traceability allows questions such as "what ends will be affected if this means is affected?" Additionally, it is usually possible to answer the question "how well does this means satisfy the end we want?" One of the most popular tools to trace (and then compare) product features (i.e., means) to customer requirements (i.e., ends) is the House of Quality (HoQ) [54]. Numbers are assigned (e.g., 1-9) to each means-end relationship based on the strength that the means contributes to the end. A drawback to the HoQ is that the numerical score values used to measure the strength of the contribution are subjective (e.g., strong, medium, weak). Additionally, since the HoQ is constructed using a 2D grid, only two dimensions can be compared in the same grid, i.e., requirements can be related to software goals, but if those software goals are to be related to stakeholder goals or business goals, then additional grids will be required for each extra dimension. If these dimensions are not explored (e.g., if the software project goals are not abstracted to business goals), then the goals that the alternative solutions will be evaluated against may be incorrect (e.g., solution specific or aiming to solve the wrong problem). Despite the drawbacks to using grids, they are argued to be the best means of visually displaying traceability for large numbers of traced entities [55], since they avoid diagrammatic "spaghetti", and perhaps most importantly, they visualise the lack of traceability with empty cells (e.g., means which do not contribute to an end).

To complement Planguage, Gilb proposes an approach called impact estimation [51], which estimates the impact of alternative system options (i.e., means) against a set of requirements (i.e., ends) using a 2d grid. This approach is very similar to the approach used by Van Lamsweerde to evaluate alternative design options [12], as previously discussed in subsection IV.B.i, and as such, it shares the same problems for application to our problem. The main contribution (related to our problem) of the impact estimation method is that it allows the practitioner to represent their confidence (interval [0,1]) in their prediction of the effect a means has on an end.

## V. PROPOSED APPROACH

We propose that GRL goal graphs can be used to demonstrate strategic alignment by linking requirements as tasks (where the task is to implement the requirement) and business objectives as hard goals (where the hard goal brings about some business benefit) with contribution links (where the requirement is the means to the objective's end). The requirements should be abstracted (asking "why?") until they link to business objectives. Business objectives then link to higher objectives, until the business strategy is represented.

### A. Constructing the Goal Graph

Before looking at how software requirements and business objectives can be connected with goal graphs, we must first explain how we represent the individual concepts.

We define business objectives using an adaption of the GQM+Strategies formalisation template [56], as in Table 1. Requiring a description of a goal using a metrics template encourages more descriptive goal models, e.g., "improve component lifespan" is defined rather than "improve engine".

TABLE 1: EXAMPLE GQM+STRATEGIES FORMALISATION

| | | |
|---|---|---|
| Activity | Reduced | |
| Object | GT-BU Fabricated Structures (FS) | |
| Focus | Average Manufacturing Lead Time | |
| Magnitude | Target:      3 months [reduction]<br>Threshold:  2 months [reduction] | As-Is: 6 months |
| Scale | Average time in months required to have FS parts manufactured from the inception of a new engine | |
| Timeframe | 1 year after system deployment | |
| Scope | Gas Turbine Components X,Y & Z | |
| Author | John Smith (Component Engineer, GT-BU) | |

Our modifications to the textual template attempt to improve integration with visual GRL diagrams through:

1. The addition of the most important concept [47] from Planguage - the scale, which specifies exactly what is to be measured, and the unit of measure.
2. The addition of scale qualifiers [51] to better describe the magnitude, e.g., "threshold" separates acceptable from unacceptable [39]. When we refer to the magnitude of an objective, we refer to the target.
3. The specification of the objective's activity attribute in the past tense, since objectives represent desired outcomes rather than an activities.
4. The removal of the constraints and relations fields - these can be expressed diagrammatically with obstacles and links (e.g., dependencies), respectively.
5. The addition of the author field so that newly proposed objectives can be identified and traced.

For our reference implementation, we use the Volere template to define the attributes of a requirement, primarily because it requires a fit criterion for testing the satisfaction of a requirement. Similarly, an objective can be considered satisfied when the specified magnitude is achieved within the specified timeframe (since benefits are not realised instantly).

Figure 1 shows an example diagram produced following the approach to explore and visualise the strategic alignment of three high-level software requirements.

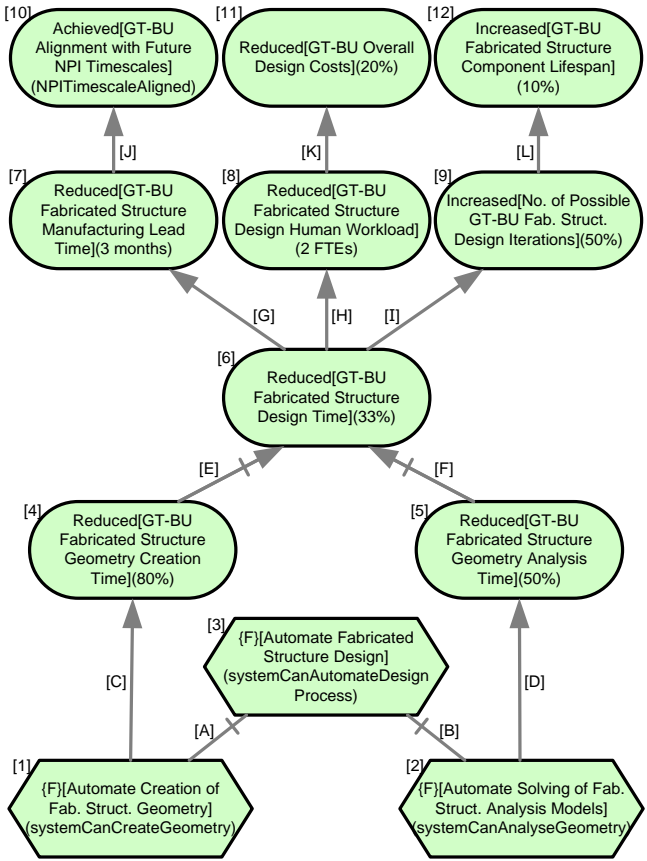

Figure 1: Example Strategic Alignment Diagram using GRL

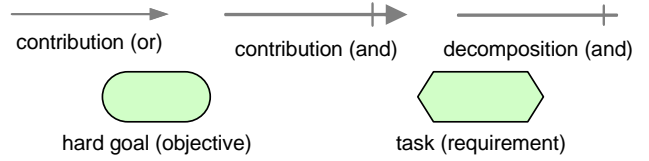

Figure 2: Key for Figure 1 - GRL Elements Used

We represent software requirements as GRL tasks (i.e., the task of implementing the requirement) using the naming syntax: "{F/NF}[Requirement](Fit Criterion)", where "F/NF" is either Functional or Non-Functional, "Requirement" is a short headline version of the requirement description, and "Fit Criterion" is the short-hand version of the metric used to test the requirement's satisfaction. In order to visualise the objectives (specified by the GQM+Strategies template) in a goal graph, we use GRL hard goals with the naming syntax: "Activity[Object Focus](Magnitude)".

Soft goal elements (e.g., goals and visions from the BMM) are not defined in the goal graph for the purpose of demonstrating strategic alignment. This is because their satisfaction criterion is undefined and thus immeasurable. Therefore, it is nonsensical to consider that a requirement may either partially or completely satisfy a goal or a vision. However, since objectives exist to quantify goals, and since goals exist to amplify the vision [26], non-weighted traceability between an objective and its goals (and their related vision) should be maintained for posterity and for impact analysis.

A contribution link between a requirement and an objective specifies that the requirement's satisfaction will achieve some satisfaction of the objective. The extent of the satisfaction is defined by the contribution score specified by the link, and is defined in terms of the objective's scale (thus making contribution scores testable). A link between two objectives is similar, except that the satisfaction of an objective is measured by its magnitude (target) rather than by a fit criterion. If the contributions of the child elements additively amount to meet the parent element's specified magnitude, then the model suggests that the parent element will be satisfied.

An "OR" contribution specifies that if there are multiple "OR" links, a decision has to be made about which should be satisfied. An "AND" contribution specifies that all "AND" links are required for the objective to be satisfied. The contribution links (E & F) are of the "AND" type, since both objectives (4 & 5) are required if objective (6) is to be satisfied. Decomposition links can be used to refine a requirement into more specific requirements, much like SysML's hierarchy link stereotype [22]. The high-level software requirement (3) is decomposed to two lower level requirements (1 & 2) to represent the hierarchy of requirement abstraction. The decomposed requirements (1 & 2) then link to objectives (4 & 5) with contribution links in order to represent what those requirements hope to achieve. The decomposition of requirements continues until the lowest level of requirements are represented. For example, requirement (2) is decomposed to specify which type of analysis should be automated (e.g., structural integrity, cost, etc.). Then, these decompositions contribute to more specific objectives (e.g., "reduce the average time taken to assess structural integrity").

## B. Single-Point Goal Graph Quantification

Both requirements and objectives have target levels of satisfaction (i.e., fit criteria and magnitudes) prescribed. This target level is a single point of possible satisfaction, where multiple points refer to satisfaction better or worse than the target level. In Table 2 (which complements Figure 1), we show a sample of quantified contribution scores for this single point of possible satisfaction (i.e., the predicted contribution if the target level of satisfaction is achieved). Note that the numbers are now fictional due to commercial sensitivity.

TABLE 2: QUANTIFIED LINK CONTRIBUTION PREDICTIONS

| Link | [Contribution] [Activity] [Scale] | Confidence |
|---|---|---|
| C (1→4) | [80%] [Reduction] in [Geometry Creation Time] | 1 |
| D (2→5) | [50%] [Reduction] in [Geometry Analysis Time] | 0.75 |
| E (4→6) | [20%] [Reduction] in [Time Required to Design] | 1 |
| F (5→6) | [13%] [Reduction] in [Time Required to Design] | 0.75 |
| G (6→7) | [3 months] [Reduction] in [Manufacturing Lead Time] | 0.75 |
| H (6→8) | [2 FTEs] [Reduction] in [Human Workload] | 1 |

The quantified contribution for link (C) tells us that objective (4) will be satisfied if requirement (1) is satisfied, since objective (4)'s required magnitude of reduction (80%) will be contributed by the complete satisfaction of requirement (1). Note that where percentages are used as contribution scores on links, this does not infer that a certain percentage of the objective's magnitude will be achieved (in this case, 80% of 80%). Instead, the focus of the objective (e.g., geometry creation time) will be affected by that percentage in the context of the activity (e.g., a reduction by at least 80%). Contribution links between pairs of objectives are read in the same way; link (E) specifies that the satisfaction of objective (4), determined by its magnitude (target) attribute, will lead to some contribution (at least 20%) toward objective (6).

This abstraction of objectives to higher level objectives allows the benefits to be expressed in terms of high-level business objectives. This is done in order to disambiguate the predicted business value by placing the quantifications into context (i.e., a large saving from a small cost may be less than a small saving from a large cost). It must be noted that a contribution link should represent causation rather than correlation, and thus care should be taken to separate the two as far as possible (guidance on this can be found in [14]).

## C. Multi-Point Goal Graph Quantification

Our approach so far represents the contribution that objectiveX makes to objectiveY when objectiveX's magnitude is completely satisfied (objectiveX is interchangeable with requirementX in this statement). However, it is likely that objectives and requirements will only be partially satisfied, i.e., their required magnitude will likely not be fully achieved. Thus, pessimistic, realistic, and optimistic views (i.e., multiple points of goal satisfaction) of strategic alignment are not currently possible. Also, it is not possible to understand the pareto optimality of software requirements (e.g., where most of the benefit is achieved and where diminishing returns starts to occur). Additionally, the potential for benefit caused by a software feature is finite, e.g., a reduction in Full Time Employees (FTEs) can be gained by task automation – up to a point. Furthermore, conflated goal contribution links whose polarity is mixed can remain hidden until multi-point contribution is modelled. Checking if the relationship between two goals is hump or U-shaped (i.e., not monotonic) will indicate that the causal pathway is more complex than is modelled, and thus the goal graph should be expanded. This separation of causal pathways is advocated both in utility theory for systems engineering in [57], and business system dynamics in [14] - which gives the example: the relationship between "increase pressure to finish work", positively, and then negatively contributes to the goal "increase employee output", as fatigue eventually overcomes motivation.

In order to understand the effects of partial satisfaction on the chain of goals, it is important to know the contribution objectiveX makes to objectiveY at various levels of objectiveX's satisfaction. This is represented by defining a *table function* [14], i.e., pairs of objectiveX and objectiveY values, together with a chosen interpolation method (linear, step-after, cardinal, monotone, etc.). Table functions are used since analytic (i.e., algebraic) functions are difficult to design, experiment with, and communicate to stakeholders when used to model non-linear relationships [14]. For simple linear relationships (e.g., converting between units on an infinite scale), algebraic functions will likely be quicker to define. Table functions should span the worst to best-case range for an objective. If the value of objectiveX is outside the function's domain, i.e., objectiveX's value is extreme, then ideally the table function should be updated, since the worst or best-case points are no longer representative. Otherwise, values lower or higher than the function's domain could be mapped to the function's minimum and maximum values. Alternatively, the slope of the last two points could support extrapolation.

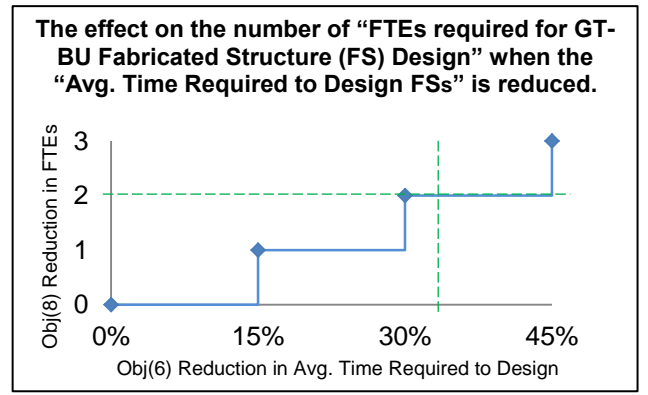

Figure 3: Link (H)'s Quantitative Contribution Relationship

To illustrate a multi-point quantified contribution link, Figure 3 visualises link (H), which is comprised of four pairs of values, and, in this case, step-after interpolation to represent integer increments (in other settings, linear interpolation and rounding may be more fitting). In this contribution link, extreme values of objective (6) are mapped to the minimum and maximum data point specified by the table function, in order to represent finite benefit realisation. Improvements to the reusability and robustness of the relationship, currently in the form of Y = *f*(X), could be made by normalising the function such that the input and output of the function are dimensionless, i.e., independent of the unit of measurement used (e.g., to define time or human resource usage). Guidance on constructing non-linear functions can be found in [14].

The visualisation appropriate to depict a contribution link depends on the type of numerical data (i.e., discrete or continuous) used by an objective's scale or a requirement's fit criterion. For functional requirements, a bar chart should be used, since they have two states (i.e., implemented or not), whereas non-functional requirements should be represented using line (xy) charts, since they have infinite states of satisfaction. Note that the green lines on the axes represent the magnitudes (targets) required by the respective objectives, as specified by the goal formalisation template (as in Table 1).

*D. Describing Confidence in Quantifications*

Contribution links in goal graphs are predictions of a causality relationship between two goals. Epistemological uncertainty (caused by a lack of knowledge) about a contribution link therefore must exist to some degree, since we cannot have perfect knowledge about future events. Before we look to describe confidence in goal contribution links, we must first distinguish *certainty* from *confidence*.

When predicting unknown quantities, uncertainty refers to beliefs about possible values for the unknown quantity, while confidence refers to the belief that a given predicted value is correct [58]. For example with reference to contribution link (H), uncertainty represents the range of possible values of FTE reduction (e.g., an interval [0, maxWorkloadInFTEs]) that could reasonably occur given a reduction in design time of 33%, i.e., the satisfaction of objective (8). In terms of Figure 3, uncertainty would affect the thickness of the line (i.e., lack of precision) used to represent the causation. Confidence, on the other hand, represents the belief that the chosen prediction (e.g., 2 FTEs) is the correct one. Thus a stakeholder's confidence is influenced by the salient factors that they believe to affect the correctness of their prediction, while a stakeholder's uncertainty is influenced by the number of different prediction options that could be correct [58].

In this paper, we focus on confidence, since empirical studies have shown that while practitioners can judge which of their predictions are more uncertain, they find quantifying the uncertainty interval difficult [59]. However, if a stakeholder were reluctant to provide a single value to quantify the contribution, the contribution could be specified in more uncertain terms, such as: 2 ± 1 FTEs for link (H), i.e., an *interval estimate* [60]. It is important that the range is restricted as far as possible to avoid ambiguity in the contribution description, since the utility of a prediction is diminished by imprecision. If the range of uncertainty is wide, it should be expressed with a Probability Density Function (PDF) to show which points in the range are more or less certain [13]. For a single-point contribution relationship (where it is assumed that the target of the first objective will be met), a PDF would describe the distribution of belief over the range of possible values for objectiveY, given a specific value of objectiveX (i.e., objectiveX's target). However, many values of objectiveX (i.e., the x-axis in Figure 3) are possible, leading to many possible PDFs to describe - reducing the usability of the approach. Thus instead, stakeholders should be encouraged to specify a single value of objectiveY that they are the most confident of, i.e., a *point estimate* [60], as in Table 2. Aleatoric uncertainty (not caused by a lack of knowledge) also exists, in that a requirement's contribution to an objective (i.e., its benefit) depends on the system's environment (i.e., context or scenario of use) [47]. This could be represented by specyifying contribution scores for each of these environments, as we exemplify in [41] (usage profiles). In this paper, we describe the average contribution made over all usage profiles (i.e., considering all likely types/contexts of use).

In the decision analysis field, it is well recognised that representing confidence is essential in determining optimal decisions, especially where a choice has to be made between two options which seem to provide similar benefits [9]. Furthermore, the description of confidence will indicate the risk that the modelled strategic alignment may not occur in practice. The confidence level representation concept we adopt is similar to that used by Gilb for impact estimation [51], so, in Table 3, we enumerate confidence levels using a similar scale. Mapping textual descriptions to confidence values (interval [0,1]) allows stakeholders to more easily select a value based on the quality of the supporting evidence (i.e., salient factors).

TABLE 3: CONFIDENCE LEVEL ENUMERATIONS

| Confidence | Description |
|---|---|
| 0.25 | Poor credibility, no supporting evidence or calculations, high doubt about capability |
| 0.5 | Average credibility, no evidence but reliable calculations, some doubt about capability |
| 0.75 | Great credibility, reliable secondary sources of evidence, small doubt about capability |
| 1 | Perfect credibility, multiple primary sources of evidence, no doubt about capability |

Basic confidence adjustment can be performed by multiplying contributions by their associated confidence level so that users are reminded of the impact confidence has on predictions, as in [51]. For example, when confidence levels are taken into consideration in Table 2, the satisfaction of requirement (1) still leads to the full satisfaction of objective (4). However, when confidence levels are considered for links (E & F), the satisfaction of objective (6) is in doubt, since (20*1) + (13*0.75) is less than the 33% required by the objective's magnitude attribute. Adjusting contributions to account for confidence in this way is similar to calculating the expected value of a random variable. However, since the mapping between the textual statements and the numbering in Table 3 (adapted from [51]) is not grounded by evidence or heuristics, a contribution score which is adjusted for confidence using them should not be treated as an expected value, but rather as an indication of the effects of confidence. If we wanted to better approximate the expected value, a number based on probability should be used to represent confidence [58], i.e., the answer to such a question: if the re-quirement were implemented a large number of times, what percentage of those times would the stated contribution be contributed? Formulating an answer to such a question depends on the experiences of the stakeholders in implementing similar requirements in similar projects in a similar environment. Similarity in this sense is difficult to achieve, since there are many socio-technical variables that can affect the benefits realised by a software project or a particular feature.

Additional confidence levels could be associated to the user's predictions to represent how qualified that user is at predicting contribution scores. For example, someone who has implemented similar systems should be able to provide more accurate predictions than someone who has not. The accuracy of a person's previous predictions (i.e., their credibility) could also be considered in order to improve the reliability of the predictions (i.e., calibrated confidence levels).

### E. Describing the Utility of a Goal's Satisfaction

One important value consideration is so far, untreated: "what is the benefit in achieving a root goal to various degrees of satisfaction?" i.e., business objectives that do not contribute to other business objectives, such as objective (12). Root objectives exist when the business has not defined any objectives higher than the objective, and where it would not make sense for them to have done so. To address this, we map various levels of a root goal's satisfaction to degrees of *utility* [9], whereby various levels of "goodness" can be achieved. For example, referring to objective (12), various levels of component lifespan improvement map to utility values (interval [0,1]). This allows the representation of non-linear relationships between component lifespan improvement and the associated benefit; perhaps after a 60% improvement on the average component's lifespan, there is no more benefit to be gained since the engine would be retired before the component would fail. Thus, the utility of a 60% improvement would peak at 1. The concept of utility is both subjective and specific to the stakeholder who assigned it. However, capturing it will explain the criticality of a root goal's satisfaction criterion, and differences in utility assignment between stakeholders will be made apparent for conflict resolution before the requirement is implemented.

Note that the maximum utility of some goal satisfaction is defined in isolation from other goals. That is to say that the maximum utility value (i.e., 1) should be defined for each root goal, and then weights can be assigned to those root goals to determine the relative utility of some goal satisfaction, in the context of the system-to-be as a whole. This is done in order to decide on the relative importance of root goals, as in [61]. Pairwise comparison or balance beam diagrams can be used to decide on, and refine the weights [57].

### F. Describing & Monitoring As-Is Values for Goals

Wherever the magnitude attribute of an objective (and related contribution scores) is/are specified a percentage, it is especially important that the objective's as-is value is described. Otherwise, it is not possible to later verify that the magnitude (i.e., change) has happened. These values can then be recorded over time in order to evaluate the system (validation) and provide a benchmark for future improvement. Furthermore, prescribed goal satisfaction levels and predicted goal contribution levels, in current and future projects, can then be made more realistic through a feedback mechanism.

### G. Describing Assumptions or Necessary Goals

When a contribution link is made between two goals, there may be an assumption that some other necessary action will occur which enables the contribution. For example, Figure 1's contribution link (K) is underpinned by the assumption that design costs can be reduced through employee time reduction. While this may seem trivial to highlight, the actual cost reducing mechanism (perhaps redundancy) may be a thorny issue, and should be communicated as early as possible for conflict resolution. To describe this assumption, a new task could be added as a decomposition of objective (11), since GRL's decomposition links represent necessity (while contribution links represent sufficiency and polarity, i.e., +ive or -ive). Assumptions made in the calculation of contribution scores should also be made explicit, e.g., in link (H), that an FTE is 40 productive hours per week. This could be represented with a GRL belief node connected to link (H).

### H. Intended Context of Use

We suggest that this approach should be performed after the high-level requirements have been elicited, so that resources are not wasted eliciting lower level requirements that do not align well to business strategy. It is especially important that the strategic alignment of solution oriented requirements (i.e., those specified for the machine [38]) is explored, since they do not explain the problem to be solved.

It is important to note that software engineers and business analysts may not know the objectives (or the goals and visions, for that matter) at different levels of the business (i.e., the project, the business unit, the department, the overall business, etc.). Therefore, managers should work with stakeholders to define the business objectives before the requirements can be abstracted toward them. Indeed, it is likely that some software requirements will be abstracted toward business objectives that were not previously elicited.

Practitioners may find difficulty in quantifying the benefits of requirements, especially for non-functional requirements where the subject may be intangible, however, proxy indicators can usually be identified relatively easily (e.g., by polling customers to quantify customer satisfaction using a likert scale) [51]. Furthermore, where stakeholders are unable to quantitatively explain the causal relationship between a requirement and higher goals (either in the description of the numeric scales or in their values), the risk that the contribution may not occur as expected will have been indicated.

While we have focused on the benefits of software requirements, both sides of the value equation need to be considered (i.e., costs). Effort estimation models such as COCOMO [62] would be useful in predicting the development cost of a requirement. The effort required by the end users to use an implemented requirement should also be considered.

### I. Tool Support

Tool support (GoalViz) has been developed (free to download at [63]) to support the approach through:

- Input support for requirements, objectives, and contribution data (with graphical function input).
- Automatic diagram drawing to focus the user on the approach and data rather than the graph layout.
- Project libraries to facilitate learning about the contributions predicted in previous projects to improve future quantification of confidence assignment.
- Automatic evaluation and summarisation of chains of links to enable efficient understanding.
- What-if analysis allowing comparison of outcomes for different inputs where there is some uncertainty.

## VI. CONCLUSION AND FUTURE WORK

The presented approach facilitates the disambiguation of a requirement's business value through the enrichment of contribution links in a goal graph. The approach is *descriptive* (a goal is abstracted to describe the underlying problem), *prescriptive* (a certain amount of goal satisfaction is required), and *predictive* (a quantitative goal contribution score predicts how much of the prescription will be achieved by the means). This paper's unique contribution includes:

1. We have argued that the strategic alignment of software requirements depends on the contribution they make to business objectives, and since they are quantitative in nature, reasoning about the contribution made toward them should also be quantitative.
2. We have argued that since strategic alignment is based on predictions of benefit, confidence (and sometimes uncertainty) should be made explicit.
3. We have shown that the non-linear dynamics of goal contribution links can be explored as quantitative causation relationships (defined with table functions) through more than one level of goal abstraction, in order to understand the effects of partial requirement satisfaction on high-level goals.

Future work will evaluate this approach (and those related to it) against required capabilities elicited from our industrial partners. We have outlined two case studies within different industrial settings, such that the benefits and challenges can be evaluated in the context of a range of domains. Feedback resulting from the evaluations in industry will be used to improve the approach and the tool. Planned investigations into optimising the utility and the usability of the approach include empirically evaluating the:

1. extent to which stakeholder utility functions for a goal's satisfaction can be aggregated to represent the preferences and uncertainty of a collective;
2. optimal representation of uncertainty, confidence, and credibility in the causal relationships;
3. optimal way (especially with regards to reusability) to specify the causal relationship between two variables (i.e., gauge variables, KPIs, or goal satisfaction levels), e.g., with causal loop diagrams and dimensionless table functions using Vensim® [14], or specifying "business formulae" as in [40];
4. optimal way to maintain traceability between requirements and design artefacts, perhaps through SysML Requirements [22] and a GRL UML profile.

ACKNOWLEDGEMENTS

The authors wish to thank Ralph Boyce from Rolls-Royce for his valued participation and feedback in this project, and also Rolls-Royce for granting permission to publish.

REFERENCES

[1] R. Ellis-Braithwaite, R. Lock, R. Dawson, and B. Haque, "Modelling the Strategic Alignment of Software Requirements using Goal Graphs," in *7th International Conference on Software Engineering Advances*, 2012, pp. 524–529.
[2] J. Thorp, *The Information Paradox: Realizing the Business Benefits of Information Technology*. McGraw-Hill, 1999.
[3] A. Aurum and C. Wohlin, "Aligning requirements with business objectives: A framework for requirements engineering decisions," in *Requirements Engineering Decision Support Workshop*, 2005.
[4] Frederick P. Brooks, *The Mythical Man Month and Other Essays on Software Engineering*, 2nd ed. Addison Wesley, 1995.
[5] J Luftman, "Assessing business-IT alignment maturity," *Commun Assoc Inf Syst 4, Article 4*, 2000.
[6] J. Luftman and T. Ben-Zvi, "Key Issues for IT Executives 2011: Cautious Optimism in Uncertain Economic Times," *MIS Quarterly Executive*, vol. 10, no. 4, 2011.
[7] S. Singh and C. Woo, "Investigating business-IT alignment through multi-disciplinary goal concepts," *Requirements Engineering*, vol. 14, no. 3, pp. 177–207, 2009.
[8] The Standish Group, "CHAOS Summary for 2010," 2010.
[9] R. A. Howard, "The foundations of decision analysis revisited," *Advances in decision analysis: From foundations to applications*, pp. 32–56, 2007.
[10] A. Van Lamsweerde, "Requirements engineering: from craft to discipline," in *16th ACM SIGSOFT FSE*, 2008, pp. 238–249.
[11] International Telecommunication Union, "Z.151 : User requirements notation (URN) - Language definition." [Online]. Available: http://www.itu.int/rec/T-REC-Z.151/en. [Accessed: 02-Jul-2012].
[12] A. Van Lamsweerde, "Reasoning about alternative requirements options," *Conceptual Modeling: Foundations and Applications*, pp. 380–397, 2009.
[13] E. Letier and A. Van Lamsweerde, "Reasoning about partial goal satisfaction for requirements and design engineering," in *ACM SIGSOFT SEN*, 2004, vol. 29, pp. 53–62.
[14] J. Sterman, *Business Dynamics: Systems Thinking and Modeling for a Complex World*. McGraw-Hill/Irwin, 2000.
[15] S. Liaskos, R. Jalman, and J. Aranda, "On eliciting contribution measures in goal models," in *20th IEEE RE*, 2012, pp. 221–230.
[16] B. Boehm, "Value-based software engineering: Seven key elements and ethical considerations," *Value-Based Software Engineering*, pp. 109–132, 2006.
[17] S. Roberson and J. Robertson, "Volere: Requirements Specifcation Template." The Atlantic Systems Guild, 2012.
[18] M. Jackson and P. Zave, "Four Dark Corners of Requirements Engineering," *ACM TOSEM*, vol. 6, no. 1, 1997.
[19] D. T. Ross and K. E. Schoman Jr, "Structured analysis for requirements definition," *IEEE TSE*, no. 1, pp. 6–15, 1977.
[20] R. Stevens, *Systems Engineering: Coping With Complexity*. Pearson Education, 1998.
[21] The Institute of Electrical and Electronics Engineers, *IEEE Std 830-1998: IEEE Recommended Practice for Software Requirements Specifications*. IEEE-SA Standards Board, 1998.

[22] M. S. Soares and J. Vrancken, "Model-driven user requirements specification using SysML," *Journal of Software*, vol. 3, no. 6, pp. 57–68, 2008.

[23] A. Goknil, I. Kurtev, and K. van den Berg, "A metamodeling approach for reasoning about requirements," in *Model Driven Architecture–Foundations and Applications*, 2008, pp. 310–325.

[24] T. Ōno, *Toyota Production System: Beyond Large-Scale Production*. Productivity Press, 1988.

[25] A. Van Lamsweerde, "Goal-oriented requirements engineering: A guided tour," *5th IEEE RE*, pp. 249–262, 2001.

[26] Object Management Group, "BMM 1.1." [Online]. Available: http://www.omg.org/spec/BMM/1.1/. [Accessed: 16-Mar-2012].

[27] B. Boehm, "Value-Based Software Engineering: Overview and Agenda," USC-CSE-2005-504, Feb. 2005.

[28] J. M. Akkermans and J. Gordijn, "Value-based requirements engineering: exploring innovative e-commerce ideas," *Requirements Engineering*, vol. 8, no. 2, pp. 114–134, Jul. 2003.

[29] A. Herrmann and M. Daneva, "Requirements Prioritization Based on Benefit and Cost Prediction: An Agenda for Future Research," in *16th IEEE RE*, 2008, pp. 125–134.

[30] J. Gordijn, E. Yu, and B. van der Raadt, "E-service design using i* and e3value modeling," *IEEE Software*, vol. 23, no. 3, pp. 26–33, 2006.

[31] A. Dardenne, A. Van Lamsweerde, and S. Fickas, "Goal-directed requirements acquisition," *Science of Computer Programming*, vol. 20, no. 1–2, pp. 3–50, Apr. 1993.

[32] E. Yu, "Modelling Strategic Relationships for Process Rengineering," University of Toronto, 1995.

[33] D. Amyot, S. Ghanavati, J. Horkoff, G. Mussbacher, L. Peyton, and E. Yu, "Evaluating goal models within the goal-oriented requirement language," *International Journal of Intelligent Systems*, vol. 25, no. 8, pp. 841–877, 2010.

[34] L. Liu, "GRL Ontology," *University of Toronto Computer Science*. [Online]. Available: http://www.cs.toronto.edu/km/GRL/grl_syntax.html. [Accessed: 14-Aug-2012].

[35] J. Horkoff and E. Yu, "Comparison and evaluation of goal-oriented satisfaction analysis techniques," *Requirements Eng*, pp. 1–24, 2012.

[36] J. Karlsson, "Software requirements prioritizing," in *2nd IEEE RE*, 1996, pp. 110–116.

[37] A. Babar, B. Wong, and A. Q. Gill, "An evaluation of the goal-oriented approaches for modelling strategic alignment concept," in *5th IEEE RCIS*, 2011, pp. 1–8.

[38] M. Jackson, *Software requirements & specifications: a lexicon of practice, principles, and prejudices*. ACM Press, 1995.

[39] A. Pourshahid, D. Amyot, L. Peyton, S. Ghanavati, P. Chen, M. Weiss, and A. J. Forster, "Business process management with the user requirements notation," *Electronic Commerce Research*, vol. 9, no. 4, pp. 269–316, Aug. 2009.

[40] J. Horkoff, D. Barone, L. Jiang, E. Yu, D. Amyot, A. Borgida, and J. Mylopoulos, "Strategic business modeling: representation and reasoning," *Softw Syst Model*, pp. 1–27, 2012.

[41] R. Ellis-Braithwaite, "Analysing the Assumed Benefits of Software Requirements," in *19th REFSQ, Proceedings of the Workshops and the Doctoral Symposium*, 2013.

[42] N. Maiden, J. Lockerbie, D. Randall, S. Jones, and D. Bush, "Using Satisfaction Arguments to Enhance i* Modelling of an Air Traffic Management System," in *15th IEEE RE*, 2007, pp. 49–52.

[43] R. S. Kaplan and D. P. Norton, "Linking the balanced scorecard to strategy," *California Management Review*, vol. 39, no. 1, 1996.

[44] A. Babar, D. Zowghi, and E. Chew, "Using Goals to Model Strategy Map for Business IT Alignment," in *5th BUSITAL*, 2010.

[45] S. J. Bleistein, K. Cox, J. Verner, and K. T. Phalp, "B-SCP: A requirements analysis framework for validating strategic alignment of organizational IT based on strategy, context, and process," *Information and Software Technology*, vol. 48, no. 9, pp. 846–868, Sep. 2006.

[46] T. Gilb and A. Cockburn, "Point/Counterpoint," *IEEE Software*, vol. 25, no. 2, pp. 64–67, Apr. 2008.

[47] N. Maiden, "Improve Your Requirements: Quantify Them," *Software, IEEE*, vol. 23, no. 6, pp. 68–69, Dec. 2006.

[48] T. Gilb, "What's Wrong With Agile Methods? Some Principles and Values to Encourage Quantification," in *Agile Software Development Quality Assurance*, Information Science Reference, 2007.

[49] N. Juristo, A. M. Moreno, and A. Silva, "Is the European industry moving toward solving requirements engineering problems?," *IEEE Software*, vol. 19, no. 6, pp. 70–77, 2002.

[50] J. Yen and W. A. Tiao, "A systematic tradeoff analysis for conflicting imprecise requirements," in *3rd IEEE RE*, 1997, pp. 87–96.

[51] T. Gilb, *Competitive Engineering: A Handbook For Systems Engineering, Requirements Engineering, and Software Engineering Using Planguage*. Butterworth-Heinemann Ltd, 2005.

[52] V. Basili, J. Heidrich, M. Lindvall, J. Münch, M. Regardie, D. Rombach, C. Seaman, and A. Trendowicz, "Bridging the gap between business strategy and software development," in *28th ICIS*, 2007, pp. 1–16.

[53] A. Trendowicz, J. Heidrich, and K. Shintani, "Aligning Software Projects with Business Objectives," in *21st IWSM-MENSURA*, 2011, pp. 142–150.

[54] J. R. Hauser and D. Clausing, "The house of quality," *Harvard Business Review*, pp. 63–73, 1988.

[55] D. L. Moody, P. Heymans, and R. Matulevičius, "Visual syntax does matter: improving the cognitive effectiveness of the i* visual notation," *Requirements Engineering*, vol. 15, no. 2, pp. 141–175, Jun. 2010.

[56] V. Mandić, V. Basili, L. Harjumaa, M. Oivo, and J. Markkula, "Utilizing GQM+Strategies for business value analysis: an approach for evaluating business goals," in *ACM-IEEE ESEM*, New York, NY, USA, 2010, pp. 20:1–20:10.

[57] D. M. Buede, *The Engineering Design of Systems: Models and Methods*. John Wiley & Sons, 2011.

[58] D. K. Peterson and G. F. Pitz, "Confidence, uncertainty, and the use of information," *Journal of Experimental Psychology: Learning, Memory, and Cognition*, vol. 14, no. 1, pp. 85–92, 1988.

[59] A. Herrmann, "REFSQ 2011 Live Experiment about Risk-Based Requirements Prioritization: The Influence of Wording and Metrics," in *17th REFSQ, Proceedings of the Empirical Track*, 2011.

[60] R A. Kent, "Estimation," in *Data Construction and Data Analysis for Survey Research*, 2001.

[61] W. Heaven, D. Sykes, J. Magee, and J. Kramer, "A case study in goal-driven architectural adaptation," *Software Engineering for Self-Adaptive Systems*, pp. 109–127, 2009.

[62] B. Boehm, B. Clark, E. Horowitz, C. Westland, R. Madachy, and R. Selby, "Cost models for future software life cycle processes: COCOMO 2.0," *Annals of Software Engineering*, vol. 1, no. 1, pp. 57–94, 1995.

[63] R. Ellis-Braithwaite, "GoalViz Tool." [Online]. Available: http://www.goalviz.info/IJAS/index.html. [Accessed: 10-Jan-2013].